\documentclass{article}
\usepackage{enumerate}
\usepackage{amssymb}
\usepackage{amsthm}
\usepackage[intlimits,reqno]{amsmath}
\usepackage[mathcal]{eucal}
\usepackage{verbatim}
\include{diagxy}

\usepackage{enumerate}
\usepackage{verbatim}
\newcommand{\norm}[1]{\left\Vert#1\right\Vert}
\renewcommand{\sc}[2]{\langle #1|#2 \rangle}

\newcommand{\mn}[1]{\langle #1 \rangle}
\newcommand{\proj}[2]{|#1 \rangle \langle #2|}

\newcommand{\bra}[1]{\langle #1 |}
\newcommand{\cat}[1]{| #1 \rangle}
\newcommand{\abs}[1]{\left\vert#1\right\vert}

\newcommand{\R}{\mathbb R}
\newcommand{\D}{\mathbb D}

\newcommand{\No}{\mathbb{N}\cup\{0\}}

\newcommand{\Z}{\mathbb Z}
\newcommand{\C}{\mathbb C}
\newcommand{\N}{\mathbb N}
\newcommand{\I}{\mathbb I}
\renewcommand{\P}{\mathcal{P}}
\newcommand{\CP}{\mathbb{CP}}

\newcommand{\K}{\mathcal{K}}

\newcommand{\T}{\mathcal{T}}

\renewcommand{\H}{\mathcal{H}}

\newcommand{\Li}{L^\infty(\mathcal{H})}

\renewcommand{\O}{\mathcal{O}}
\newcommand{\M}{\mathfrak{M}}

\renewcommand{\to}{\rightarrow}
\newcommand{\tto}{\longrightarrow}

\renewcommand{\leq}{\leqslant}
\renewcommand{\geq}{\geqslant}
\renewcommand{\phi}{\varphi}

\DeclareMathOperator{\Tr}{Tr}

\DeclareMathOperator{\id}{id}

\DeclareMathOperator{\im}{im}

\DeclareMathOperator{\Ad}{Ad}

\newcommand{\be}{\begin{equation}}
\newcommand{\ee}{\end{equation}}
\newcommand{\bse}{\begin{subequations}}
\newcommand{\ese}{\end{subequations}}
\newcommand{\ben}{\begin{enumerate}}
\newcommand{\een}{\end{enumerate}}
\newcommand{\bit}{\begin{itemize}}
\newcommand{\eit}{\end{itemize}}

\newtheorem{thm}{Theorem}[section]
\newtheorem{cor}[thm]{Corollary}

\newtheorem{prop}[thm]{Proposition}
\theoremstyle{definition}

\theoremstyle{remark}

\numberwithin{equation}{section}
\newcommand{\prf}[1]{{\bf Proof:} #1 \qed}

\numberwithin{equation}{section}
\usepackage[cp1250]{inputenc}
\renewcommand{\M}{\mathbb{M}}

\begin{document}
\ifx\href\undefined\else\hypersetup{linktocpage=true}\fi 

\bigskip
\begin{center}
{\bf \Large Quantum Complex Minkowski Space}\\
\bigskip
{\large G. Jakimowicz, A. Odzijewicz  }\\ \bigskip
 Institute of Mathematics\\
 University  in Bialystok\\
 ul. Lipowa 41, PL-15-424 Bialystok, Poland\\
 e-mail: gjakim@alpha.uwb.edu.pl, aodzijew@uwb.edu.pl
\end{center}

\begin{abstract}
The complex Minkowski phase space has the physical interpretation of the phase space of the scalar massive
conformal particle.
The aim of the paper is the construction and investigation of the quantum complex Minkowski space.
\end{abstract}

\tableofcontents

\section{Introduction}

Extending the Poincar\'e group by dilation and acceleration transformations, one obtains the conformal
group $SU(2,2)/\Z_4$, which is
the symmetry group of the conformal structure of compactified Minkowski space-time $M$, where
$\Z_4=\{i^k \id : k=0,1,2,3\}$ is the centralizer of $SU(2,2)$.
According to the prevailing point of view $SU(2,2)/\Z_4$ is the symmetry group for physical models which describe massless fields or particles,
but has no application to the theory of massive objects. However, using the twistor
description \cite{penrose} of Minkowski space-time
and the orbit method \cite{kirillov}, the different orbits of  $SU(2,2)/\Z_4$ in the
conformally compactified complex Minkowski space $\M:=M^\C$ may be considered to be the classical phase spaces of massless and massive scalar conformal particles, antiparticles and tachyons, see \cite{Opierwsza,Oconf}.

The motivation for various attempts to construct models of non-commutative Minkowski space-time is the belief that
this is the proper way to avoid divergences in quantum field theory \cite{madore}.
Here, on the other hand, our aim is to quantize the classical phase space $\M^{++}\subset\M$ of the massive particle by replacing it by the
Toeplitz-like operator $C^*$-algebra $\mathcal M^{++}$. To this end we first quantize the classical states of the massive
scalar conformal particle by constructing the coherent state map $\K:\M^{++}\to\CP(\H)$ of $\M^{++}$ into
the complex projective Hilbert space $\CP(\H)$, i.e. the space of the pure states of the system. In the next step we define the
Banach algebra $\overline{\mathcal P}^{++}$ of annihilation operators as the ones having the coherent states $\K(m)$, $m\in\M^{++}$,
as eigenvectors. Finally, the quantum phase space $\mathcal M^{++}$ will be the $C^*$-algebra generated by $\overline{\mathcal P}^{++}$.

Let us remark that application of the above method of quantization to the case of $\R^{2N}$ phase space leads to
the Heisenberg-Weyl algebra. In our construction the conformal group and $\mathcal M^{++}$ are related in exactly the same way as
are the Heisenberg group and the Heisenberg-Weyl algebra.

The conformally compactified
Minkowski space $M$ can be reconstructed from $\mathcal M^{++}$ as the \v Silov boundary of the interior of the spectrum
of the commutative Banach algebra $\mathcal P^{++}$. It can also be considered in the framework of Kostant-Souriau quantization
as the $SU(2,2)/\Z_4$-invariant configuration space for the phase space $T^*M$. Similarly, if we consider the holomorphic model \cite{Oconf},
see Section 2 and Section 3, the classical conformal phase space $\M^{++}$ has the interpretation of the configuration space
constructed by the $SU(2,2)/\Z_4$-invariant K\"ahler polarization. In \cite{Oconf} a model of the classical field theory
on $\M^{++}$ was proposed.
This paper is an effort, developing the results presented in \cite{Oker}, to construct a
quantum description of the conformal massive particle, see Section 4.
In Section 5  the physical interpretation of the
quantum phase space $\mathcal M^{++}$ is discussed.

\section{Complex Minkowski space as the phase space of the conformal scalar massive particle}

Following \cite{Opierwsza,Oconfsymp,Oconf}, we present the twistor description of phase spaces of the conformal
scalar massive particles.
Let us recall that twistor space $\mathbf T$ is
$\mathbb C^{4}$ equipped with a Hermitian form  $\eta$ of signature $(++--)$.
The symmetry group of $\mathbf T$ is the group $\rm{SU}(2,2)$, where  $g\in \rm{SU}(2,2)$ iff $g^{\dag}\eta g=\eta$ and $\det g=1$.

In relativistic mechanics the elementary phase
spaces are given by the coadjoint orbits of the Poincar\'e group, see \cite{souriau}, which are parametrized in this case by mass, spin,
and signature of the energy of the relativistic particle. Similarly, elementary
phase spaces for conformal group 
one identifies with its coadjoint orbits. Since conformal Lie algebra $su(2,2)$ is simple
we will identify its dual $su(2,2)^*$ with $su(2,2)$ using Cartan-Killing form:
\be \langle X,Y\rangle =\frac12\Tr(XY) ,\ee
where $X,Y\in\textrm{su(2,2)}$. Thus
the coadjoint representation $\Ad^*_g:su(2,2)^*\to su(2,2)^*$ is identified with the adjoint one
\be \Ad_{g}X=gXg^{-1},\ee
where  $g\in \rm{SU}(2,2)$. For the complete description and
physical interpretation of $\Ad^{\star}(\rm{SU}(2,2))$-orbits see \cite{Otwist,Oconfsymp}.

One defines the compactified complex Minkowski space $\mathbb M$ as
the Grassmannian of 2-dimensional complex vector subspaces $w\subset \mathbf T$ of the twistor space and $\rm{SU}(2,2)$ acts on $\mathbb M$
by
\be \label{action} \sigma_{g}: w\rightarrow gw. \ee
The Grassmannian $\mathbb M$ splits  into the orbits $\M^{kl}$ indexed by
the signatures of the restricted Hermitian forms $sign  \;\eta |_z=(k,l)$, where $k,l=+,-,0$.

The orbit
$\mathbb M^{00}$ consisting of subspaces isotropic with respect to $\eta$ is the conformal compactification $M$ of real Minkowski space
and $\mathbb M$ is the complexification
of  $M=\mathbb M^{00}$.

The cotangent bundle $T^{\star}\mathbb M^{00}\to\mathbb M^{00}$ is isomorphic
with the vector bundle
$\{(x,X)\in\mathbb M^{00}\times su(2,2): \im X\subset x\subset \ker X\}=:\mathbb N\xrightarrow{pr_1} \mathbb M^{00}$,
where  $pr_1$
is the projection on the first component of the product $\mathbb M^{00}\times su(2,2)$.
The vector bundle isomorphism $T^{\ast}\mathbb M^{00}\cong\mathbb N$ is defined by the following sequence
$T_{x}^{\ast}\mathbb M^{00} \cong(su(2,2)\diagup su(2,2)_x)^{\ast}
\cong\{X\in su(2,2):\Tr YX=0\;\;\forall Y\in su(2,2)_x\}\cong\{X\in
su(2,2):\im X\subset x\subset \ker X\}= pr_{1}^{-1}(x)$ of the vector space isomorphisms.

There exists a conformal structure on $\mathbb N$ defined by the cones
$C_x:=\{X\in pr_{1}^{-1}(x) : \dim_{\mathbb R}\im X\leqslant1\}\subset pr_{1}^{-1}(x)\cong T_{x}^{\ast}\mathbb M^{00}$,
$x\in\mathbb M^{00}$. 
This conformal structure is invariant with respect to the action of $\rm{SU}(2,2)$ on $\mathbb N$ defined by
\be\label{conf-action}\alpha_{g}:(x,X)\mapsto(gx,gXg^{-1})\ee
for $g\in \rm{SU}(2,2)$.

 The 8-dimensional orbits of the action \eqref{conf-action} are: the bundle $\mathbb N^{++}\to\mathbb M^{00}$ of upper halves of the interiors of the cones,
 the bundle $\mathbb N^{--}\to\mathbb M^{00}$ of bottom halves of the interiors of the cones and
 the bundle $\mathbb N^{+-}\to\mathbb M^{00}$ of exteriors of the cones.

 Similarly, the action \eqref{action} of $\rm{SU}(2,2)$
 on $\M$
 generates three 8-dimensional orbits:  $\mathbb
 M^{++}$, $\mathbb M^{--}$ and $\mathbb M^{+-}$.

 One has maps $J_0:\tilde{\mathbb N}\to su(2,2)$ and $J_\lambda:\tilde{\mathbb M}\to su(2,2)$ of
$\tilde{\mathbb N} :=\mathbb N^{++}\cup\mathbb N^{--}\cup\mathbb N^{+-}$ and
$\tilde{\mathbb M} :=\mathbb M^{++}\cup\mathbb M^{--}\cup\mathbb M^{+-}$ into $su(2,2)$ defined by:
 \begin{align}\label{j0} J_{0}(x,X)&:=X \\
\label{jl} J_{\lambda}(w)&:=i\lambda(\pi_{w}-\pi_{w^{\bot}}),\end{align}
where $\perp : \tilde{\mathbb M} \mapsto \tilde{\mathbb M}$ is the involution, which maps $w\in\tilde{\mathbb M}$ on its
 orthogonal complement $w^\perp$ (with respect to the twistor forms $\eta$) and $\pi_{w}:\mathbf T\mapsto \mathbf T$ and $\pi_{w^{\perp}}:\mathbf T\mapsto \mathbf T$ are
 the projections defined by the decomposition $\mathbf T=w\oplus w^\perp$.

 The maps $J_0$ and $J_\lambda$ are equivariant with respect to the actions $\alpha$ and $\sigma$ respectively and $\Ad$-action of the conformal group.
Thus $J_0$ maps $\mathbb N^{++}$, $\mathbb N^{+-}$, $\mathbb N^{--}$ on
 the 8-dimensional nilpotent $\Ad$-orbits and $J_\lambda$ maps $\mathbb M^{++},\mathbb M^{++},\mathbb M^{++}$
 on the 8-dimensional
 simple $\Ad$-orbits which consist of $X\in su(2,2)$ with eigenvalues $i\lambda$ and $-i\lambda$.
 Using the Kirillov construction \cite{kirillov} we obtain the conformally invariant symplectic form $\omega_{0}$ on
 $\tilde{\mathbb N}$ (identical with the canonical symplectic form of $T^{\ast}\mathbb M^{00}$)
and  the conformally invariant
 K\"{a}hler form $\omega_{\lambda}$ on $\tilde{\mathbb M}$.
 So  $(\tilde{\mathbb N},
 \omega_{0})$ and $(\tilde{\mathbb M},\omega_{\lambda})$ are 8-dimensional conformal symplectic manifolds with momentum maps given by \eqref{j0},\eqref{jl}.

 In order to show that $\tilde{\mathbb N}$ and $\tilde{\mathbb M}$ have a physical interpretation
 of the phase spaces of the conformal scalar massive
 particles, let us take the coordinate description of the presented models.
We fix an element $\infty \in \mathbb M^{00}$, called
point at infinity. One defines the Minkowski space $\mathbb M^{00}_{\infty}$ as the affine space of elements $w \in \mathbb
M^{00}$ which are transversal to $\infty$, i.e. $w\oplus \infty =\mathbf T$. The elements $w\in\mathbb M^{00}$
which intersect with $\infty$ in more than one-dimension, i.e. $\dim_\C(w\cap\infty) \geq 1$, form a cone $C_{\infty}$
at infinity, so
$$\mathbb M^{00}=\mathbb M^{00}_{\infty}\cup C_{\infty} \cong\mathbb S^{1}\times\mathbb S^{3}.$$
The cones $C_{x}=\{x^{'}\in\mathbb M^{00}:\dim_{\mathbb C}(x\cap x')\geq1\}$ define a conformal structure on
$\mathbb M^{00}$, invariant with respect to the conformal group action given by \eqref{action}.

The Poincar\'e group $\mathbf P_{\infty}$ extended
by the dilations is defined as the stabilizer $(\rm{SU}(2,2)/\Z_4)_{\infty}$
 of the element $\infty$. The intersections of the stabilizers $\mathbf
(\rm{SU}(2,2)/\Z_4)_{\infty}\cap(\rm{SU}(2,2)/\Z_4)_{0}$, where $0\in\mathbb M^{00}_{\infty}$ is the origin of the inertial
coordinates system, is the Lorentz group extended by dilations.
One defines the Lorentz group $\mathbf L_{0,\infty}$
and the group of dilations $\mathbf D_{0,\infty}$ respectively as the commutator and the centralizer of $(\rm{SU}(2,2)/\Z_4)_{\infty}\cap
(\rm{SU}(2,2)/\Z_4)_{0}$ respectively. Finally, the group of Minkowski space translations $\mathbf T_\infty$ consists of the elements $\exp X$, where
$X\in su(2,2)$ satisfies $\im X\subset\infty\subset \ker X$, while the elements $\exp X$ fulfilling $\im X\subset 0 \subset \ker X$, define
the commutative subgroup $\mathbf A_0$ of four-accelerations.

Let us assume in the following that
\be \eta=i\begin{pmatrix}
  0  & \sigma_{0} \\
  -\sigma_{0} & 0 \\
\end{pmatrix},
\infty=\left\{\left(%
\begin{array}{c}
  \zeta \\
  0 \\
\end{array}
\right):\zeta\in\C^2\right\},   \mathbf{0}=\left\{\left(%
\begin{array}{c}
  0 \\
  \zeta \\
\end{array}%
\right):\zeta\in\C^2\right\}, \ee
where we use the $2\times 2$ matrix representation with Pauli basis: $$\sigma_{0}=\left(%
\begin{array}{@{}cc@{}}
  1 & 0 \\
  0 & 1 \\
\end{array}%
\right),\quad \sigma_{1}=\left(%
\begin{array}{@{}cc@{}}
  0 & 1 \\
  1 & 0 \\
\end{array}%
\right), \quad \sigma_{2}=\left(%
\begin{array}{@{}cc@{}}
  0 & i \\
  -i & 0 \\
\end{array}%
\right), \quad\sigma_{3}=\left(%
\begin{array}{@{}cc@{}}
  1 & 0 \\
  0 & -1 \\
\end{array}%
\right)$$
 in $Mat_{2\times2}(\C)$. This choice of $\eta ,\infty ,\mathbf{0}$ gives us the decomposition
\be su(2,2)=\mathcal T_{\infty}\oplus
\mathcal L_{0,\infty}\oplus \mathcal D_{0,\infty}\oplus \mathcal A_{0}\ee where the subalgebras of 4-translations, Lorentz,
dilations and 4-accelerations are
given respectively by \begin{subequations}\begin{align}\label{t} \mathcal T_{\infty}&=\{\left(%
\begin{array}{@{}cc@{}}
  0 & T \\
  0 & 0 \\
\end{array}%
\right) :T=T^{\dagger}\in Mat_{2\times 2}(\mathbb{C})\textrm{ and }T=t^\mu\sigma_\mu\} \\
\label{l} \mathcal L_{0,\infty}&=\{\left(%
\begin{array}{@{}cc@{}}
  L & 0 \\
0 & -L^\dag \\
\end{array}%
\right): \Tr L=0\textrm{ and }L\in Mat_{2\times 2}(\mathbb{C})\} \\
 \label{d} \mathcal D_{0,\infty}&=\{d\left(%
\begin{array}{@{}cc@{}}
  \sigma_{0} & 0 \\
  0 & -\sigma_{0} \\
\end{array}%
\right):d\in\mathbb{R}\} \\
\label{a} \mathcal A_{0}&=\{\left(%
\begin{array}{@{}cc@{}}
  0 & 0 \\
  C & 0 \\
\end{array}%
\right):C=C^{\dag}\in Mat_{2\times 2}(\mathbb{C})\textrm{ and }C=c^\mu\sigma_\mu\}
\end{align}
\end{subequations}
The basis of $su(2,2)^{\ast}\cong su(2,2)$ dual to the one defined by Pauli
matrices in the Lie subalgebras  $\mathcal T_\infty$, $\mathcal L_{0,\infty}$, $\mathcal D_{0,\infty}$, $\mathcal A_0$
 is
\begin{subequations}\label{dual-pauli}\begin{align}
\label{bp} \mathcal T_\infty^*\ni \mathcal P^{\ast}_{\mu}&=\left(%
\begin{array}{@{}cc@{}}
  0 & 0 \\
  \sigma_{\mu} & 0 \\
\end{array}%
\right) \\
 \label{bl}\mathcal L_{0,\infty}^*\ni \mathcal L^{\ast}_{kl}&=\frac{1}{2}\epsilon_{klm}\left(%
\begin{array}{@{}cc@{}}
  \sigma_{m} & 0 \\
  0 & \sigma_{m} \\
\end{array}%
\right)\qquad
\mathcal L_{0,\infty}^*\ni\mathcal L_{0k}^\ast&=\frac{1}{2}\left(%
\begin{array}{@{}cc@{}}
  \sigma_{k} & 0 \\
  0 & -\sigma_{k} \\
\end{array}%
\right)\\
 \label{bd} \mathcal D_{0,\infty}^*\ni\mathcal D^{\ast}&=\frac{1}{2}\left(%
\begin{array}{@{}cc@{}}
  \sigma_{0} & 0 \\
  0 & -\sigma_{0} \\
\end{array}%
\right)\\
\label{ba}\mathcal A_0^*\ni\mathcal A_{\nu}^{\ast}&=\left(%
\begin{array}{@{}cc@{}}
  0 & \sigma_{\nu} \\
  0 & 0 \\
\end{array}%
\right)
\end{align}
\end{subequations}
One has the matrix coordinate  map
\be\mathbb{M_{\infty}}\ni w\mapsto W\in Mat_{2\times 2}(\C)\ee
defined by \be \label{mat-coord}w=\{\left(%
\begin{array}{c}
  W\zeta \\
  \zeta \\
\end{array}%
\right):\zeta\in \C^{2}\}\ee
and $w=x\in\mathbb M^{00}_{\infty}$ iff $W=W^{\dag}=X.$ The element $(x,\mathcal X)\in pr_{1}^{-1}(\mathbb M^{00}_{\infty})$ is parametrized by \be (x,\mathcal X)\mapsto (X,\begin{bmatrix}
  XS & -XSX  \\
  S & -SX\\
\end{bmatrix}),\ee
where $X,S\in H(2)$ and $H(2)$ is the vector space of $2\times 2$ Hermitian matrices.

The momentum maps \eqref{j0} and \eqref{jl} in the above defined coordinates are given by
\be J_{0}(X,S)=\begin{bmatrix}
  XS & -XSX \\
  S & -SX \\
\end{bmatrix}\ee \be J_{\lambda}(W)=i\lambda\begin{bmatrix}
  (W+W^{\dag})(W-W^{\dag})^{-1} & -2W(W-W^{\dag})^{-1}W^\dag \\
  2(W-W^{\dag})^{-1} & -\sigma_{0}-2(W-W^{\dag})^{-1}W^{\dag} \\
\end{bmatrix}. \ee
By decomposing $J_0(X,S)$ in the basis \eqref{dual-pauli}
 $J_{0}(X,S)=p_\mu\mathcal P^*_\mu+ m_{\mu\nu}\mathcal L^*_{\mu\nu} + a_\mu\mathcal A^*_\nu + d\mathcal D^*$
we obtain the expressions
\begin{align} \label{p}p_{\mu}&=s_{\mu}\\
\label{m}m_{\mu\nu}&=x_{\mu}p_{\nu}-x_{\nu}p_{\mu}\\
\label{dd}d&=x^{\mu}p_{\mu}\\
\label{aa}a_{\mu}&=-2(x^{\nu}p_{\nu})x_{\mu}+x^{2}p_{\mu}
\end{align}
for the four-momentum $p_{\mu}$, relativistic angular
momentum $m_{\mu\nu}$, dilation $d$ and four-acceleration $a_{\nu}$ respectively, where $S=s^{\mu}\sigma_{\mu}$, $X=x^{\mu}\sigma_{\mu}$.

In the coordinates $x_\mu$, $p_\mu=s_\mu$
the symplectic form $\omega_{0}$ assumes the canonical form
\be \omega_{0}=dx^{\mu}\wedge
dp_{\mu}.\ee

Similarly, from
 $J_{\lambda}(W)=p_\mu\mathcal P^*_\mu+ m_{\mu\nu}\mathcal L^*_{\mu\nu} + a_\mu\mathcal A^*_\nu + d\mathcal D^*$
 we obtain
\begin{align} \label{pp}p^{\nu}&=\lambda\frac{y^{\nu}}{y^{2}}\\
m_{\mu\nu}&=x_{\mu}p_{\nu}-p_{\nu}x_{\mu}\\
d&=x^{\mu}p_{\mu}\\
\label{aaa}a_{\mu}&=-2(x^{\nu}p_{\nu})x_{\mu}+x^{2}p_{\mu}-\frac{\lambda^2}{p^2}p_\mu,\end{align}
where the real coordinates $x_\mu,y_\mu$ on $\tilde{\mathbb M}$ are defined by  $x^{\nu}+iy^{\nu}=w^{\nu}:=\frac12 \Tr(W\sigma_\nu)$.

The coordinate description of $\omega_\lambda$ is the following
\be \omega_{\lambda}=i\lambda\frac{\partial^{2}}{\partial w^{\mu}\partial
\bar{w}^{\nu}}\log(w-\bar{w})^2 dw^\mu\wedge d\bar w^\nu=dx^{\nu}\wedge dp_{\nu}.\ee

Concluding, one has two models $(\mathbb{N},\omega_{0})$ and $(\mathbb{M},\omega_\lambda)$
of the massive scalar conformal
particle. Using the canonical coordinates $(x^{\mu},p_{\nu})$ common for both models we obtain that
\begin{enumerate}[(i)]
\item the element $(x,\mathcal X)\in\mathbb{N^{++}}$ ($w\in\mathbb M^{++})$ iff $p^{0}>0$ and $(p^{0})^{2}-\overrightarrow{p}^{2}>0$, i.e.
it describes the state of a conformal scalar massive particle;
\item the element $(x,\mathcal X)\in\mathbb{N^{--}}$ ($w\in\mathbb M^{--})$ iff $p^{0}<0$ and $(p^{0})^{2}-\overrightarrow{p}^{2}>0$, i.e.
it describes the state of a conformal scalar massive anti-particle;
\item the element $(x,\mathcal X)\in\mathbb{N^{+-}}$ ($w\in\mathbb M^{+-})$ iff $(p^{0})^{2}-\overrightarrow{p}^{2}<0$, i.e.
it describes the state of a conformal scalar tachyon.
\end{enumerate}
The orbits $\mathbb N^{0+}$ ($\mathbb M^{0+}$) and  $\mathbb N^{0-}$ ($\mathbb M^{0-}$) describe the states of massless particles and anti-particles
but this case will not be discussed further.

Two above presented models do not differ if one considers them on the level
of relativistic mechanics, since both of them
behave towards Poincar\'e  transformations in the same way. The difference  appears if one considers
the four-acceleration transformations parametrized by $C=c^\mu\sigma_\mu$, which in canonical coordinates $X=x^\mu\sigma_\mu$, $P=p^\mu\sigma_\mu$
are
\be\tilde{X}=X(CX+\sigma_{0})^{-1},\ee\be\tilde{P}=(CX+\sigma_{0})P(XC+\sigma_{0})\ee
for the  standard model $\tilde{\mathbb N}$ and
 \be \label{tildeX}\tilde{X}=[XP+i\lambda\sigma_{0}-i\lambda(XC-i\lambda P^{-1}
C+\sigma_{0})](CXP+i\lambda C+P)^{-1}\ee
\be \label{tildeP}\tilde{P}=(CX+\sigma_{0})P(XC+\sigma_{0})+\lambda^{2}CP^{-1}C\ee
for the holomorphic model $\tilde{\mathbb M}$. We see from \eqref{tildeP} that in the holomorphic model (opposite to the
standard one) the four-momentum $P=p^\mu\sigma_\mu$ transforms in a non-linear way. This fact implies
a lot of important physical consequences, e.g.
the conformal scalar massive particle cannot be localized in the space-time in conformally invariant way.
From \eqref{aaa}, \eqref{tildeX}, \eqref{tildeP} it follows that the
holomorphic model corresponds to the nilpotent one when $\lambda\rightarrow 0$.

\section{Conformally invariant quantum K\"{a}hler polarization}

In this section we shall make the first step in the direction to construct quantum conformal phase space.
Since the case of the antiparticle can be transformed by the charge conjugation map to the particle one, see \cite{Oconf},
and the tachyon case is less
interesting from physical point of view, we will work only with the phase space $\mathbb{M^{++}}$ of the
conformal scalar massive particle.

The phase space $T^{\ast}\mathbb M^{00}$
has the real conformally invariant 
polarization defined by the leaves of its
cotangent bundle  structure. In canonical coordinates this polarization is spanned by the vector
fields $\{\frac{\partial}{\partial p^{\nu}}\}_{\nu =0,...3}$. For the holomorphic phase space $\tilde{\mathbb{M}}$
the conformally invariant polarization is K\"{a}hler and in the complex coordinate it is spanned by
$(\frac{\partial}{\partial\bar{w^{\mu}}})_{\mu=0,...3}$. The reason is that $\rm{SU}(2,2)/\Z_4$ acts on
$\tilde{\mathbb{M}}$ by biholomorphism. For $ g^{-1}=\left(%
\begin{array}{@{}cc@{}}
  A & B \\
  C & D \\
\end{array}%
\right)\in \rm{SU}(2,2)$ and $w\in \mathbb{M^{++}}$
 one has
\be \sigma_{g}W=(AW+B)(CW+D)^{-1},\ee
where $W\in Mat_{2\times
2}(\mathbb{C})$ is the matrix holomorphic coordinate of $w\in\mathbb{M^{++}}$. Using complex matrix coordinates \eqref{mat-coord} one identifies
$\mathbb{M^{++}}$ with the future tube
\be \mathbb{T}:=\{W\in Mat_{2\times 2} :\im W> 0\}. \ee
Applying the Caley transform
\be \label{caley}Z=(W-iE)(W+iE)^{-1}   ,W=i(Z+E)(Z-E)^{-1}\ee
we map $\mathbb{T}$ on the symmetric domain
\be \mathbb{D}:=\{Z\in Mat_{2\times 2}(\mathbb{C}):  E-Z^{\ast}Z>0\}. \ee
Let us remark here that the coordinates $Z\in\mathbb D$ correspond to the
diagonal representation of the twistor form $\eta=\left(%
\begin{array}{@{}cc@{}}
  \sigma_{0} & 0 \\
  0 & -\sigma_{0} \\
\end{array}%
\right)$. Below we use both systems of coordinates.

In order to quantize $\mathbb{M^{++}}$ we will use the method of coherent state map investigated in \cite{Ocoh}.
For the other construction of noncommutative manifolds by using coherent state method see also \cite{grosse}. The essence of
this method consists in replacing the classical state $m\in\mathbb{M^{++}}$  by the quantum pure state, which means,
that one defines the map $\mathcal
K_{\lambda}:\mathbb{M^{++}}\mapsto\mathbb{C}\mathbb{P}(\mathcal H)$ from the classical phase space $\mathbb{M^{++}}$ into the complex projective separable Hilbert space
$\mathbb{C}\mathbb{P}(\mathcal H)$. We will call $\mathcal K_\lambda$ coherent state map and in our case we will postulate that it has the following properties:

\begin{enumerate}[(i)]
\item $\mathcal K_\lambda$ is consistent with the conformal symmetry, i.e. there  exists an unitary irreducible representation
$\mathbf{U_{\lambda}}:\rm{SU}(2,2)\mapsto \operatorname{Aut}\mathcal H$ with respect to which the coherent state map is equivariant:

\be\label{Ab}\bfig \square<700,500>[\mathbb M^{++}`\mathbb{CP}(\H)`\mathbb M^{++}`\mathbb{CP}(\H);\mathcal K_\lambda`\sigma_g`[\mathbf U_{\lambda}{(g)]}\;\;\forall g\in \rm{SU}(2,2)`\mathcal K_\lambda] \efig \ee


\item $\mathcal K_\lambda$ is consistent with the holomorphic polarization $(\frac{\partial}{\partial\bar{w^{\mu}}}_{\mu=0,...3})$. This denotes
 that $\mathcal K_{\lambda}$ is a holomorphic map.

\item $\mathcal K_\lambda$ is symplectic, i.e.
\be \mathcal K^{\ast}_{\lambda}\omega_{FS}=\omega_{\lambda},\ee
where $\omega_{FS}$ is
Fubini-Study form on $\mathbb{C}\mathbb{P}(\mathcal H)$. The projective Hilbert space is considered here as K\"{a}hler
manifold (thus symplectic manifold).
This condition one needs for the consistence of classical dynamics with quantum dynamics.
\end{enumerate}

The coherent state map $\mathcal K_{\lambda}:\mathbb{M^{++}}\mapsto\mathbb{C}\mathbb{P}(\mathcal H)$ fulfilling
the properties postulated above one obtains by the applying of the representation theory, see \cite{perelomov,pere}.
We skip here the technical considerations and present only the final result. Let
\be\label{baza}\left\{\left|\begin{array}{@{}cc@{}}
  j & m \\
  j_{1} & j_{2} \\
\end{array}\right\rangle\right\},\ee
where $m,2j\in\mathbb{N}\cup\{0\}$ and $-j\leq j_{1},j_{2}\leq j$, denote an orthonormal basis
in $\mathcal H$, i.e.
\be \left\langle \begin{array}{@{}cc@{}}
  j & m \\
  j_{1} & j_{2} \\
\end{array}\right.\left|\begin{array}{@{}cc@{}}
  j' & m' \\
  j_{1}' & j_{2}' \\
\end{array}\right\rangle
=\delta_{jj'}\delta_{mm'}\delta_{j_1j_1'}\delta_{j_2j_2'}.\ee
Then the map $K_{\lambda}:\mathbb{M^{++}}\cong\mathbb{D}\mapsto\mathcal H$ given by
\be\label{coh}
K_\lambda: Z \to |Z;\lambda\rangle:=\sum_{j,m,j_{1},j_{2}}\Delta^{jm}_{j_{1}j_{2}}(Z)\left|\begin{array}{@{}cc@{}}
  j & m \\
  j_{1}& j_{2} \\
\end{array}\right\rangle,\ee
where
\be \Delta^{jm}_{j_{1}j_{2}}(Z):=(N^{\lambda}_{jm})^{-1}(\det Z)^{m}\sqrt{\frac{(j+j_{1})!(j-j_{1})!}{(j+j_{2})!(j-j_{2})!}}\times\ee
$$\times\!\!\!\sum_{\substack{S\geq \max\{0,j_{1}+j_{2}\}\\ S\leq \min\{j+j_{1},j+j_{2}\}}}
\!\!\!\!\left(%
\begin{array}{@{}c@{}}
  j+j_{2} \\
  S \\
\end{array}%
\right)\left(%
\begin{array}{@{}c@{}}
  j-j_{2}\\
  S-j_{1}-j_{2} \\
\end{array}%
\right)z_{11}^{S}z_{12}^{j+j_{1}-S}z_{21}^{j+j_{2}-S}z_{22}^{S-j_{1}-j_{2}}$$
and
\be
N^{\lambda}_{jm}:=(\lambda-1)(\lambda-2)^{2}(\lambda-3)\frac{\Gamma(\lambda-2)\Gamma(\lambda-3)m!(m+2j+1)!}{(2j+1)!\Gamma(m+\lambda-1)\Gamma(m+2j+\lambda)},\ee
defines a  coherent state map
\be[K_\lambda]=:\mathcal
K_{\lambda}:\mathbb{M^{++}}\mapsto\mathbb{C}\mathbb{P}(\mathcal H)\ee
 with the properties mentioned in
assumptions: (i), (ii), (iii). The condition (i) restricts the variability of the parameter $\lambda>3$ to integer numbers.

From now on, to simplify the notation, we will write $\cat{Z}$ instead of  $\cat{z;\lambda}$. If the dependence
on $\lambda$ is relevant we will write  $\cat{z;\lambda}$.

The projectors $\frac{|Z\rangle\langle Z|}{\langle Z| Z\rangle}$
representing the coherent
states give the resolution of the identity
\be \label{res-id}\mathbb{\mathbf{1}}=\int_{\mathbb{D}}|Z\rangle\langle
Z|   \;\;d\mu_{\lambda}(Z,Z^{\dag})\ee
with respect to the measure \be\label{me}
d\mu_{\lambda}(Z,Z^{\dag})=c_{\lambda}[\det(E-Z^{\dag}Z)]^{\lambda-4}|dZ|,\ee
where $|dZ|$ is the Lebesgue measure on
$\mathbb{D}$ and
\be c_\lambda=\pi^{-4}(\lambda-1)(\lambda-2)^{2}(\lambda-3),\ee
 which is equivalent to $\int_\mathbb D d\mu_\lambda=1$.

Hence, by the anti-linear monomorphism
\be \label{anti-I} I_{\lambda}: \mathcal
H\ni|\psi\rangle\rightarrow\langle\psi|\;\cdot\;;\lambda\rangle:=\psi(\cdot)\in\mathcal O(\mathbb{D})\ee
one identifies $\mathcal
H$ with the range of $I_{\lambda}$ in $\O(\D)$, which is equal to the Hilbert space of holomorphic functions $L^{2}\mathcal
O(\mathbb{D},d\mu_{\lambda})$ square integrable with respect to the measure (\ref{me}).

The representation
$I_{\lambda}\circ{U}_\lambda\circ I_{\lambda}^{-1}$ acts on $L^{2}\mathcal O(\mathbb{D},d\mu_{\lambda})$ by
\be \label{rep1}(I_{\lambda}\circ{U}_\lambda(g)\circ
I_{\lambda}^{-1})\psi(Z)=[\det(CZ+D)]^{-\lambda}\psi(\sigma_{g}(Z)),\ee
i.e. it is a discrete series representation of $\rm{SU}(2,2)$ and acts on the coherent states by
\be\label{rep}
{U}_{\lambda}(g)|Z\rangle=[\det(CZ+D)]^{-\lambda}|\sigma_{g}(Z)\rangle,\ee
where
$g^{-1}=\left(%
\begin{array}{@{}cc@{}}
  A & B \\
C & D \\
\end{array}%
\right)\in \rm{SU}(2,2)$, see \cite{graev,ruhl}.

The fifteen physical quantities $p_{\nu}$, $m_{\mu\nu}$, $d$ and $a_{\nu}$, $\mu$, $\nu=0,1,2,3$, which characterize the scalar massive
conformal particle form the conformal Lie algebra $su(2,2)$ with respect to the Poisson bracket
\be\label{poiss}\{f,g\}_\lambda(\bar w,w)=
\frac{i}{2\lambda}\left((w-\bar w)^2\eta^{\mu\nu}-2(w^\mu-\bar w^\mu)(w^\nu-\bar w^\nu)\right)\left( \frac{\partial f}{\partial w^\mu}\frac{\partial g}{\partial \bar w^\nu}-\frac{\partial g}{\partial w^\mu} \frac{\partial f}{\partial \bar w^\nu}\right)\ee
defined by the
symplectic form $\omega_{\lambda}$. Each one of them defines a Hamiltonian flow $\sigma_{g(t)}$ on $\mathbb{M^{++}}$
realized by the corresponding one-parameter subgroup $g(t)$, $t\in\R$, of $\rm{SU}(2,2)$. By the equivariance condition \eqref{Ab} this Hamiltonian
flow $\sigma_{g(t)}$ is quantized to the Hamiltonian flow on $\CP(\H)$ given by the one-parameter subgroup $\mathbf U_\lambda(g(t))$ of
representation \eqref{rep}.
The
generators of these one-parameter subgroups are realized in $L^{2}\mathcal O(\mathbb{T},d\mu_{\lambda})$ as follows:
\begin{align}\label{hat-p}\hat{p_{\mu}}&=-i\frac{\partial}{\partial w^{\mu}}\\
\hat{m_{\mu\nu}}&=-i(w_{\mu}\frac{\partial}{\partial
w^{\nu}}-w_{\nu}\frac{\partial}{\partial w^{\mu}})\\
\hat{d}&=-2iw^{\mu}\frac{\partial}{\partial w^{\mu}}-2i\lambda\\
\label{hat-a}\hat{a_{\nu}}&=-iw^{2}(\delta^{\beta}_{\nu}-2w_{\nu}w^{\beta})\frac{\partial}{\partial w^{\beta}}+2i\lambda
w_{\nu},\end{align}
see \cite{Oker}.
They are quantized versions of their classical counterparts
given by \eqref{p}-\eqref{aa}. The measure $d\mu_{\lambda}$ in the future tube
representation is given by
\be \label{miara} d\mu_{\lambda}(W,W^{\dag})=2^{-4}[\det(W-W^{\dag})]^{\lambda-4}|dW|.\ee
It was shown in \cite{Ocoh} that the coherent state  method of
quantization is equivalent to the Kostant-Souriou geometric quantization.

Besides generators \eqref{p}-\eqref{aa} of the conformal Lie algebra $su(2,2)$ there is also reason to quantize other physically
important observables. In particular case the ones belonging to the family $\O^{++}(\D)$ consisting of complex valued smooth functions
$f:\M^{++}\to\C$ for whose there exists bounded operators $a(f)\in \Li$ such that
\be \label{z} a(f)\cat{Z}=f(Z)\cat Z\ee
for any $Z\in\D\cong\M^{++}$. Since the coherent states $\cat Z$ form a linearly dense subset of $\H$ one has
correctly defined linear map $a:\mathcal \O^{++}(\D)\to\Li$ of $\mathcal O^{++}(\D)$ in the Banach algebra of the bounded operators.

It follows immediately from \eqref{z} and the resolution of identity \eqref{res-id} that
\ben[i)]\item $\O^{++}(\D)$ is the commutative algebra and $f\in\O^{++}(\D)$ is holomorphic;
\item The map $a:\O^{++}(\D)\to\Li$ is an isometric
\be \norm{a(f)}_\infty=\norm{f}_{\sup}=\sup_{Z\in\D}\abs{f(Z)}\ee
monomorphism of algebras;
\item The image $a(\O^{++}(\D))$ is uniformly closed in $\Li$ (i.e. with respect to operator norm $\norm\cdot_\infty$).\een
Hence, $\O^{++}(\D)$ is a Banach subalgebra of the Banach algebra $H^\infty(\D)$ of functions which are holomorphic and bounded on $\D$. Let
us remark here that completeness of $H^\infty(\D)$ follows from the Weierstrass theorem, see e.g. \cite{shabat2}.

Indeed one has:
\begin{prop} The Banach algebra $\O^{++}(\D)$ is equal to $H^\infty(\D)$.\end{prop}
\prf{Since $I_\lambda(\H)=L^2\O(\D,d\mu_\lambda)$ we have $f\sc{\psi}\;\cdot\; \in I_\lambda(\H)$ for any $f\in H^\infty(\D)$.
The multiplication operator $M_f:L^2\O(\D,d\mu_\lambda)\to L^2\O(\D,d\mu_\lambda)$ is bounded. Thus there is a bounded operator
$a(f)^{\ast}:\H\to\H$ such that
\be f(Z)\sc\psi Z=\sc{a(f)^*\psi}Z\ee
for $Z\in\D$. The above shows that $a(f)=(a(f)^*)^*$ fulfills \eqref{z}.}

According to \cite{Okah}  we shall call the commutative Banach algebra $\mathcal P^{++}:=a(H^\infty(\D))$ the \emph{quantum
K\"ahler polarization} and its elements $a(f)\in \mathcal P^{++}$ the \emph{annihilation operators}.

The coordinate functions $f_{kl}(Z):=z_{kl}$, where $k,l=1,2$ belong to $H^\infty(\D)$. Therefore
$a_{kl}:=a(f_{kl})\in\mathcal P^{++}$ and their action on the basis \eqref{baza} is given by
\begin{eqnarray} a_{11}\left|\begin{array}{@{}cc@{}}
  j & m \\
  j_{1}& j_{2} \\
\end{array}\right\rangle
&=\sqrt{\frac{(j-j_{1}+1)(j-j_{2}+1)m}{(2j+1)(2j+2)(m+\lambda-2)}}\left|\begin{array}{@{}cc@{}}
  j+\frac{1}{2} & m-1 \\
  j_{1}-\frac{1}{2}& j_{2}-\frac{1}{2} \\
\end{array}\right\rangle \nonumber \\ &+\sqrt{\frac{(j+j_{1})(j+j_{2})(m+2j+1)}{(m+2j+\lambda-1)2j(2j+1)}}\left|\begin{array}{@{}cc@{}}
  j-\frac{1}{2} & m \\
  j_{1}-\frac{1}{2}& j_{2}-\frac{1}{2} \\
\end{array}\right\rangle\end{eqnarray}

\begin{eqnarray} a_{12}\left|\begin{array}{@{}cc@{}}
  j & m \\
  j_{1}& j_{2} \\
\end{array}\right\rangle
&=-\sqrt{\frac{(j-j_{1}+1)(j+j_{2}+1)m}{(2j+1)(2j+2)(m+\lambda-2)}}\left|\begin{array}{@{}cc@{}}
  j+\frac{1}{2} & m-1 \\
  j_{1}-\frac{1}{2}& j_{2}+\frac{1}{2} \\
\end{array}\right\rangle \nonumber \\
&+\sqrt{\frac{(j+j_{1})(j-j_{2})(m+2j+1)}{(m+2j+\lambda-1)2j(2j+1)}}\left|\begin{array}{@{}cc@{}}
  j-\frac{1}{2} & m \\
  j_{1}-\frac{1}{2}& j_{2}+\frac{1}{2} \\
\end{array}\right\rangle\end{eqnarray}

\begin{eqnarray} a_{21}\left|\begin{array}{@{}cc@{}}
  j & m \\
  j_{1}& j_{2} \\
\end{array}\right\rangle
&=-\sqrt{\frac{(j+j_{1}+1)(j-j_{2}+1)m}{(2j+1)(2j+2)(m+\lambda-2)}}\left|\begin{array}{@{}cc@{}}
  j+\frac{1}{2} & m-1 \\
  j_{1}+\frac{1}{2}& j_{2}-\frac{1}{2} \\
\end{array}\right\rangle \nonumber \\ &+\sqrt{\frac{(j-j_{1})(j+j_{2})(m+2j+1)}{(m+2j+\lambda-1)2j(2j+1)}}\left|\begin{array}{@{}cc@{}}
  j-\frac{1}{2} & m \\
  j_{1}+\frac{1}{2}& j_{2}-\frac{1}{2} \\
\end{array}\right\rangle\end{eqnarray}

\begin{eqnarray} a_{22}\left|\begin{array}{@{}cc@{}}
  j & m \\
  j_{1}& j_{2} \\
\end{array}\right\rangle
&=\sqrt{\frac{(j+j_{1}+1)(j+j_{2}+1)m}{(2j+1)(2j+2)(m+\lambda-2)}}\left|\begin{array}{@{}cc@{}}
  j+\frac{1}{2} & m-1 \\
  j_{1}+\frac{1}{2}& j_{2}+\frac{1}{2} \\
\end{array}\right\rangle \nonumber \\
&+\sqrt{\frac{(j-j_{1})(j-j_{2})(m+2j+1)}{(m+2j+\lambda-1)2j(2j+1)}}\left|\begin{array}{@{}cc@{}}
  j-\frac{1}{2} & m \\
  j_{1}+\frac{1}{2}& j_{2}+\frac{1}{2} \\
\end{array}\right\rangle. \end{eqnarray}
In the expressions above we put by definition $\left|\begin{array}{@{}cc@{}}
  j & m \\
  j_1& j_2 \\
\end{array}\right\rangle:=0$ if the indices do not satisfy the condition $m,2j\in\mathbb{N}\cup\{0\}$ and $-j\leq j_{1},j_{2}\leq j$.

The coordinate annihilation operators $a_{kl}, k,l=1,2$ generate Banach subalgebra $\mathcal P^{++}_{pol}$ of $\mathcal P^{++}$.
Let us denote by $Pol(\overline \D)$ the algebra of polynomials of variables $\{z_{kl}\}, k,l=1,2$ restricted to the closure
$\overline\D$ of $\D$ in $Mat_{2\times2}(\C)$.

For the following considerations let us fix the matrix notation
\be \label{A}\mathbb{A}:=\left(%
\begin{array}{@{}cc@{}}
  a_{11} & a_{12} \\
  a_{21} & a_{22} \\
\end{array}%
\right)\in\P^{++}_{pol}\otimes Mat_{2\times2}(\C),\ee
\be \mathbb A^+:=\left(\begin{array}{@{}cc@{}} a_{11}^* & a_{21}^* \\ a^*_{12} & a^*_{22} \end{array}\right)
\in \overline{\P^{++}_{pol}}\otimes Mat_{2\times 2}(\C)\ee
for the annihilation and creation operators. For example, in this notation the
property \eqref{z} assumes the form
\be \mathbb{A}|Z\rangle=Z|Z\rangle.\ee

\begin{prop}$\;$

\ben[i)]
\item $\mathcal P^{++}_{pol}$ is isometrically isomorphic to the closure $\overline{Pol(\overline \D)}$ of $Pol(\overline \D)$, i.e.
$a(f)\in \mathcal P^{++}_{pol}$ iff $f$ is continuous on $\overline\D$ and holomorphic on $\D$. The space
of maximal ideals of the $\mathcal P^{++}_{pol}$ (the spectrum) is homeomorphic to $\overline \D$.
\item $\mathcal P^{++}_{pol}$ is a semisimple Banach algebra, i.e. if $p\in \mathcal P^{++}_{pol}$ is such that for each $c\in\C$ there exists
$(1+cp)^{-1}$ then $p=0$.
\item $\mathcal P^{++}_{pol}\varsubsetneq \mathcal P^{++}$, i.e. it is proper Banach subalgebra of $\mathcal P^{++}$.
\item The vacuum state is cyclic with respect to the Banach algebra $\overline{\mathcal P^{++}_{pol}}$
\een\end{prop}
\prf{$\;$

 $i)$ For $Z,W\in\overline\D$ and $\alpha\in [0,1]$ one has
\be v^\dag(E-[\alpha
Z+(1-\alpha)W]^{\dag}[\alpha Z+(1-\alpha)W])v=\norm v^2-\|[\alpha Z+(1-\alpha)W]v\|^{2}\geqslant\ee
$$
\|v\|^{2}-\{\alpha\|Zv\|+(1-\alpha)\|Wv\|\}^{2}\geqslant\|v\|^{2}-\{\alpha\|v\|+(1-\alpha)\|v\|\}^{2}=0,$$
for each $v\in\C^2$, what gives
$\alpha Z+(1-\alpha)W\in\overline{\mathbb{D}}$. So, $\overline \D$ is convex bounded subset of $Mat_{2\times 2}(\C)$. Thus $\overline\D$
is polynomially convex and compact. By definition $\mathcal P^{++}_{pol}$ has a finite number of generators. Hence statement $i)$
is valid, see for example Chapter 7 of \cite{wermer}.

$ii)$ We recall that the radical of algebra $\mathcal{P}^{++}$ is
\be\mathcal{R}=\{b\in\mathcal{P}^{++}:(b+\lambda\mathbb I)\textrm{ is invertible for any }\lambda\neq0\}.\ee

$iii)$ To prove this it is enough to find a function $f\in H^\infty(\D)$ such that $f\notin\overline{Pol(\overline\D)}$.
For example  the function
\be f(Z)=\exp\frac{\Tr(Z+E)}{\Tr(Z-E)}\ee
has this property.

$iv)$ It is enough to check that
\be \label{jmjj}\left|\begin{array}{@{}cc@{}}
  j & m \\
  j_{1}& j_{2} \\
\end{array}\right\rangle=\Delta^{j m}_{j_1 j_2}(\mathbb A^\dag)\left|\begin{array}{@{}cc@{}}
  0 & 0 \\
  0& 0 \\
\end{array}\right\rangle\ee
and notice that operator $\Delta^{j m}_{j_1 j_2}(\mathbb A^\dag)\in\overline{\mathcal P^{++}_{pol}}$.}

We define the action of the $g\in \rm{SU}(2,2)$ on $\mathbb{A}$ by
\be \mathbf U_\lambda(g)\mathbb{A}\mathbf U_\lambda(g^{-1}):=\left(%
\begin{array}{@{}cc@{}}
  U_\lambda(g)a_{11}U_\lambda(g^{-1}) &  U_\lambda(g)a_{12}U_\lambda(g^{-1}) \\
   U_\lambda(g)a_{21}U_\lambda(g^{-1}) & U_\lambda(g)a_{22}U_\lambda(g^{-1})  \\
\end{array}%
\right),\ee
where $\rm{SU}(2,2)\ni g\to U_\lambda(g)\in\operatorname{Aut}(\H)$ is discrete series representation
defined by \eqref{rep1}. Using the above notation
we formulate the following statement.

\begin{prop}\label{prop2}One has
\ben[i)]
\item \be\sigma_{g}(\mathbb{A}):=(A\mathbb{A}+B)(C\mathbb{A}+D)^{-1}\in\P^{++}_{pol}\otimes Mat_{2\times2}(\C),\ee
\item
\be \label{UG}\mathbf U_\lambda(g)\mathbb{A}\mathbf U_\lambda(g^{-1})=\sigma_g(\mathbb A),\ee
for $g\in \rm{SU}(2,2)$.
\een
\end{prop}
\prf{$\;$

$i)$ For $g^{-1}=\left(%
\begin{array}{cc}
  A & B \\
  C & D \\
\end{array}%
\right)\in \rm{SU}(2,2)$ one has
\be DD^\dag=E+CC^\dag.\ee
So eigenvalues of $DD^\dag$ satisfy $d_1, d_2\geq 1$, which implies that
\be \norm{D^{-1}CZ}^2\leq \norm{D^{-1}C(D^{-1}C)^\dag}=\norm{D^{-1}(DD^\dag-E){D^\dag}^{-1}}=\norm{E-D^{-1}{D^\dag}^{-1}}<1\ee
for $Z\in\overline \D$. The above gives that
$(D+CZ)^{-1}=(E+D^{-1}CZ)^{-1}D^{-1}$ exists for $Z\in\overline\D$.
Since $\det(CZ+D)$ is continuous function function of $Z$ and  $\det(CZ+D)\neq 0$ for $z\in\overline \D$
there exists $\Omega\supset \overline\D$ such that $\det(CZ+D)\neq 0$ for all $z\in\Omega$. This shows that
the matrix function
\be\sigma_{g}(\mathbb{Z})=(A\mathbb{Z}+B)(C\mathbb{Z}+D)^{-1}\ee
is holomorphic on $\Omega$. So, by the Oka-Weil theorem, see \cite{wermer,shabat}, there is a sequence $\{p_n\}$ of polynomials in $z_{11}, z_{12}, z_{21}, z_{22}$ with
$p_n\to \sigma_g$ uniformly on $\overline\D$. Since $\mathcal P^{++}_{pol}\cong\overline{Pol(\overline\D)}$ one proves
$\sigma_g(\mathbb A)\in \mathcal P^{++}_{pol}\otimes Mat_{2\times 2}(\C)$.

$ii)$ Let us note that for a linearly dense set of the coherent states $\cat Z$, $Z\in\D$,
\be \mathbf U_\lambda(g)\mathbb A \mathbf U_\lambda(g^{-1}) |Z\rangle=\sigma_g(\mathbb A)|Z\rangle\ee
which gives \eqref{UG}. }

We conclude immediately from Proposition \ref{prop2}
\begin{cor}
  Banach subalgebra $\mathcal P^{++}_{pol}\subset \Li$ is invariant $U_\lambda(g)\mathcal P^{++}_{pol}U_\lambda(g^{-1})\subset \mathcal P^{++}_{pol}$,
  $g\in \rm{SU}(2,2)$ with respect to the discrete series representation.
\end{cor}

Let us make a closing remark that quantum polarization $\mathcal P^{++}$ gives holomorphic operator coordinatization
for the classical phase space $\M^{++}$ and subalgebra $\mathcal P^{++}_{pol}\subset\mathcal P^{++}$ gives the coordinatization of
$\M^{++}$ algebraic in the annihilation operators.

\section{Conformal K\"ahler quantum phase space}
The holomorphic quantum coordinatization of the classical
phase space $\mathbb M^{++}$ by the operator Banach algebra $\P^{++}$ is not sufficient from the
physical point of view. The reason is that the complete quantum
description of the scalar conformal particle also requires self-adjoint operators,
for example such as those given by \eqref{hat-p}-\eqref{hat-a}. Therefore, we are obliged to include
in our considerations the Banach algebra $\overline{\P^{++}}$ generated by the creation operators
$a^*_{kl}$, $k,l=1,2$, which by definition are conjugated counterparts of the
annihilation operators. The algebra $\overline{\P^{++}}$ gives anti-holomorphic quantum coordinatization
of $\M^{++}$. From Proposition \ref{prop2} it follows that  $\overline{\P^{++}}$ as
well as $\P^{++}$ are conformally invariant \emph{quantum K\"ahler polarizations} on $\M^{++}$.
Then, following \cite{Okah}, we shall call
the operator $C^*$-algebra $\mathcal M^{++}\subset L^\infty(\H)$ generated by $\P^{++}$
the \emph{quantum K\"ahler phase space} of the scalar conformal particle. We shall
denote by $\mathcal M^{++}_{pol}$ the proper $C^*$-subalgebra of $\mathcal M^{++}$ generated by $\mathcal P^{++}_{pol}\varsubsetneq\mathcal P^{++}$.

The relation between the quantum phase space $\mathcal M^{++}$ and its classical mechanical counterpart $\M^{++}$ is best
seen by the covariant and contravariant symbols description.

For any bounded operator $F\in\Li$ one defines the \emph{2-covariant symbol}
\be \mn{F}_{2}(Z^\dag,V):=\frac{\sc Z{FV}}{\sc ZV }.\ee
The \emph{2-contravariant symbol} $f$ is defined as an element of the space $\mathcal B_2(\D\times\D)$ of complex valued functions on $\D\times \D$ for which the integral
\be \label{contra}F=\mathcal F_\lambda(f):=c^2_\lambda\int_{\D\times\D} f(Z^\dag,V)\frac{\cat Z \sc ZV \bra V}{\sc ZZ \sc VV} d\mu(Z^\dag,Z)d\mu(V^\dag,V) \ee
exists weakly and $\mathcal F_\lambda(f)\in\Li$, where the measure $d\mu$ is defined by
\be d\mu(Z^\dag,Z)=\det(E-Z^\dag Z)^{-4}\abs{dz}.\ee
We define:
\ben[i)]
\item the associative product
\begin{align}&(f \bullet_{\lambda} g)(Z^\dag,W):=\\
 =&c_{\lambda}^{2}\int_{\D\times\D} f(Z^\dag,V)g(S^\dag,W) \frac{\sc ZV \sc VS \sc SW }{\sc ZW \sc VV \sc SS }
d\mu(V^\dag,V)d\mu(S^\dag,S)=\nonumber  \\
=&\int_{\D\times\D} f(Z^\dag,V)g(S^\dag,W)\frac{\sc ZV \sc VS \sc SW }{\sc ZW }
d\mu_{\lambda}(V^\dag,V)d\mu_{\lambda}(S^\dag,S),\nonumber \end{align}
of the 2-contravariant symbols $f,g\in\mathcal B_2(\D\times\D)$ ;
\item the seminorm
\be \norm f:=\norm {\mathcal F_\lambda(f)}_\infty\ee
and the involution
\be f^*(Z^\dag,V):=\overline{f(V,Z^\dag)}\ee
\een of the 2-contravariant symbol.
The map $\mathcal F_\lambda:\mathcal B_2(\D\times\D)\to \Li$ is an epimorphism of algebras with involution and
\be \ker \mathcal F_\lambda=\{f\in \mathcal B_2(\D\times\D): \norm f=0\}.\ee
Thus the quotient algebra $\mathcal B_2(\D\times\D)/\ker \mathcal F_\lambda$ and $\Li$ are isomorphic as $C^*$-algebras. Since each equivalence class
$[f]=f+\ker \mathcal F_\lambda$ is represented in a unique way by the 2-covariant symbol $\mn{\mathcal F_\lambda(f)}_2$, i.e. $[f]=\mn{\mathcal F_\lambda(f)}_2+\ker \mathcal F_\lambda$
and $\mn{\mathcal F_\lambda(f)}_2=\mn{\mathcal F_\lambda(g)}_2$ iff $f-g\in \ker\mathcal F_\lambda$, then the quotient vector space $\mathcal B_2(\D\times\D)/\ker \mathcal F_\lambda$ is
isomorphic with the vector space
\be \mathcal B^2(\D\times\D):=\{\mn{F}_2: F\in \Li\}\ee
of 2-covariant symbols of the bounded operators. Defining the product of the 2-covariant symbols $\mn{F}_2, \mn G_2\in \mathcal B^2(\D\times\D)$
by
\be \label{prod-cov}\mn F_2 *_{\lambda}\mn G_2(Z^\dag,V):=c_{\lambda}\int \mn F_{2} (Z^\dag,W) \mn G_{2}(W^\dag,V)
\frac{\sc ZW \sc WV }{\sc WW \sc ZV } d\mu(W^\dag,W)\ee
one obtains the structure of $C^*$-algebra on $\mathcal B^2(\D\times\D)$.

The quotient map $\mathcal B_2(\D\times\D)\to\mathcal B_2(\D\times\D)/\ker\mathcal F_\lambda$ and the isomorphism
$\mathcal B_2(\D\times\D)/\ker\mathcal F_\lambda\cong\mathcal B^2(\D\times\D)$ defines the epimorphism
\be \pi: \mathcal B_2(\D\times\D)\tto \mathcal B^2(\D\times\D)\ee
of the algebra with involution $(\mathcal B_2(\D\times\D),\bullet_\lambda)$ on the  $C^*$-algebra $(\mathcal B^2(\D\times\D),*_\lambda)$.
Similarly the inclusion map
\be\iota:\mathcal B^2(\D\times\D)\hookrightarrow\mathcal B_2(\D\times\D)\ee
is the monomorphism of $C^*$-algebra $(\mathcal B^2(\D\times\D),*_\lambda)$ to the algebra $(\mathcal B_2(\D\times\D),\bullet_\lambda)$.

In the case under consideration the coherent state map $\mathcal K_\lambda:\D\to\CP(\H)$ is holomorphic
and $\D$ is a simply connected domain. Hence one can recontruct the 2-covariant
symbol $\mn F_2$ of the bounded operator $F\in\Li$ from its \emph{Berezin covariant symbol}

\be \mn F (Z^\dag,Z):=\frac{\sc Z{FZ}}{\sc ZZ}.\ee
The reconstruction is given by the analytic continuation of $\mn F$ from the diagonal $\delta:\D\cong\Delta\hookrightarrow\D\times\D$ to the product $\D\times\D$.
As a result we obtain the linear isomorphism
\be \label{cont}c:\mathcal B(\D)\xrightarrow{\quad\sim\quad} \mathcal B^2(\D\times\D)\ee
of the vector space $\mathcal B(\D):=\{\mn{F}:F\in\Li\}$ of Berezin covariant symbols with $\mathcal B^2(\D\times\D)$.
The map \eqref{cont} is inverse to the restriction map
\be \delta^*:\mathcal B^2(\D\times\D)\ni \mn F_2 \tto \mn F_2\circ \delta\in \mathcal B(\D).\ee
Hence one also defines the product
\be \label{star0}f*_\lambda g:=\delta^*(c(f)*_\lambda c(g))\ee
of $f,g\in\mathcal B(\D)$, which is given explicitly by
\be \label{star}(f *_\lambda g)(Z^\dag,Z)=c_\lambda\int_\D f(Z^\dag,V)g(V^\dag,Z)\abs{a_\lambda(Z^\dag,V)}^2 d\mu(V^\dag,V),\ee
where
\be \label{alamb}a_\lambda(Z^\dag,V):=\frac{\sc ZV}{\sqrt{\sc ZZ \sc VV}}\ee
is the transition amplitude between the coherent states $\mathcal K_\lambda(Z)$ and $\mathcal K_\lambda(V)$.
For brevity,
by $f$ and $g$ in \eqref{star0} we denoted the Berezin covariant symbols of $F,G\in\Li$.

Let us visualize the morphisms defined above in the following diagram

\be\label{diagram}\bfig \Vtriangle(0,0)/>`<-^)`>/[(\mathcal B_{2}(\D\times\D),\bullet_\lambda)`(\Li,\circ)`(\mathcal B^2(\D\times\D),*_\lambda);\mathcal F_\lambda`\iota`\mn\cdot_2]
\morphism(350,-650)|l|/<-/<0,500>[\quad\quad(\mathcal B(\D),*_\lambda)`;\delta^*]
\morphism(550,-570)/->/<0,420>[`;c]
\morphism(-100,450)|b|/>/<350,-350>[`;\pi]
 \efig \ee

The notions of covariant and contravariant symbols were introduced by Berezin and their importance in various aspects
of quantization was shown in \cite{ber1,ber2}. The 2-contravariant and 2-covariant symbols
of Schatten class operators and bounded operators were studied in \cite{Ober}.

In the following proposition we will mention a few properties of the quantum scalar conformal phase
space $\mathcal M^{++}$ and its $C^*$-subalgebra $\mathcal M^{++}_{pol}$.

\begin{prop}$\;$ \label{prop41}

\begin{enumerate}[(i)]
\item The autorepresentation of $\mathcal M^{++}_{pol}$ in $L^\infty(\H)$ is irreducible and
$\P^{++}_{pol}\cap \overline{\P^{++}_{pol}}=\C \mathbb I$.
\item $\mathcal M^{++}_{pol}$ is weakly (strongly) dense in $L^\infty(\H)$.
\item $\mathcal M^{++}_{pol}$ contains the ideal $L^0(\H)$ of compact operators. Thus any ideal of $\mathcal M^{++}_{pol}$, which autorepresentation
in $\H$ is irreducible, also contains $L^0(\H)$.
\item $\mathcal M^{++}_{pol}$ is conformally invariant, i.e. $U_\lambda(g)\mathcal M^{++}_{pol} U_\lambda(g)^\dag\subset \mathcal M^{++}_{pol}$   for $g\in SU(2,2)$.
\item $\mathcal P^{++}_{pol}\cap L^0(\H)=\{0\}$.
\item $L^0(\H)\subsetneq Comm \mathcal M^{++}_{pol}$, where $Comm \mathcal M^{++}_{pol}$ is commutator ideal of $\mathcal M^{++}_{pol}$.
\item The statements $i)$, $ii)$, $iii)$, $v)$, and $vi)$ are valid also for $\mathcal M^{++}$ and $\mathcal P^{++}$
\end{enumerate}
\end{prop}

\prf{$\;$
\begin{enumerate}[(i)]
\item Let us denote by $P$ the orthogonal projector defined by decomposition of $\H$ on the Hilbert subspaces irreducible
with respect to $\mathcal M_{pol}^{++}$. Let us define $p\in L^2\O(\D,d\mu_\lambda)$ by
\be p(Z):=\overline{\left \langle Z\bigg|P\bigg|\begin{array}{@{}cc@{}}
  0 & 0 \\
  0& 0 \\
\end{array}\right\rangle}.\ee
Since
\be \label{P}a(f)^\dag P=Pa(f)^\dag\ee
for each $f\in Pol(\overline\D)$ then from \eqref{jmjj} and \eqref{P}one has
\be \label{I}(I\circ P\circ I^{-1})I\left(\bigg|\begin{array}{@{}cc@{}}
  j & m \\
  j_{1}& j_{2} \\
\end{array}\right\rangle\bigg)=pI\left(\bigg|\begin{array}{@{}cc@{}}
  j & m \\
  j_{1}& j_{2} \\
\end{array}\right\rangle\bigg). \ee
Since $p\in L^2\O(\D,d\mu\lambda)$ there exists a sequence of polynomials $\{p_n\}$ such that $p_n\xrightarrow[n\to\infty]{} p$
in $\norm\cdot_2$-norm. The operator $I\circ P\circ I^{-1}$ is bounded, so we obtain from \eqref{I}
\be p=(I\circ P \circ I^{-1})p=(I\circ P \circ I^{-1})\lim_{N\to\infty}p_N=\!\!\lim_{N\to\infty}(I\circ P \circ I^{-1})p_N =
\!\!\lim_{N\to\infty}pp_N.\ee
For any compact subset $K\subset \D$ one has
\be \sup_{Z\in K} \abs{\sc\psi Z}\leq C_k \norm\psi_2,\ee
where $C_k:=\sup_{Z\in K}\sqrt{\sc ZZ}$ and thus
\begin{align} 0&\leq\sup_{Z\in K} \abs{p^2(Z)-p(Z)p_N(Z)}\leq \sup_{Z\in K} \abs{p(Z)}\sup_{Z\in K}\abs{p(Z)-p_N(Z)}\leq\nonumber\\
&\leq C_k^2\norm p_2 \norm{p-p_N}_2\xrightarrow[N\to\infty]{}0.\end{align}
The above gives $p=\lim_{N\to\infty} pp_N=p^2\in L^2\O(\D,d\mu_\lambda)$. Thus $p\equiv 1$ and from \eqref{I} we obtain
that $P=\I$, what proves irreducibility of $\mathcal M^{++}_{pol}$. If $x\in\mathcal P^{++}_{pol}\cap \overline{\mathcal P^{++}_{pol}}$
then it commutes with any element of $\mathcal M^{++}_{pol}$. So $x\in\C\I$.
\item It follows from $i)$ and from the von Neumann bicommutant theorem.
\item Let us take the operator $F\in\Li$ which has finite number of nonzero matrix elements in the orthonormal basis \eqref{baza}.
Then its 2-covariant symbol is given by
\be \mn F_2(Z^\dag,V)=\!\!\!\!\!\!\sum_{\substack{(j,m,j_1,j_2)\in\Phi\\(j',m',j_1',j_2')\in \Phi}} \!\!\!\!\!\!(\det(E-Z^\dag V)^\lambda \Delta^{j m}_{j_1 j_2}(Z^\dag)
\left\langle\begin{array}{@{}cc@{}}
  j & m \\
  j_{1}& j_{2} \\
\end{array}\right|F
\left|\begin{array}{@{}cc@{}}
  j' & m' \\
  j'_{1}& j'_{2} \\
\end{array}\right\rangle
\Delta^{j' m'}_{j'_1 j'_2}(V),\ee
where $\Phi$ is a finite index set. The operator
\be \label{D}\!\!\!\!\!\!\sum_{\substack{(j,m,j_1,j_2)\in\Phi\\(j',m',j_1',j_2')\in \Phi}} \!\!\!\!\!\!\Delta^{j m}_{j_1 j_2}(\mathbb A^\dag)
\left\langle\begin{array}{@{}cc@{}}
  j & m \\
  j_{1}& j_{2} \\
\end{array}\right|F
\left|\begin{array}{@{}cc@{}}
  j' & m' \\
  j'_{1}& j'_{2} \\
\end{array}\right\rangle
\Delta^{j' m'}_{j'_1 j'_2}(\mathbb A)\ee belongs to $\mathcal M^{++}_{pol}$ and has the same 2-covariant symbol as operator $F$.
Thus we gather that $F$ is equal to \eqref{D} what implies that  $F\in\mathcal M^{++}_{pol}$.
So from the fact that $L^0(\H)\cap \mathcal M^{++}_{pol}\neq\{0\}$ and Theorem 2.4.9 in \cite{murphy} we see that $L^0(\H)\subset\mathcal M^{++}_{pol}$.

\item Since $\mathcal M^{++}_{pol}$ is generated by $\mathcal P^{++}_{pol}$, the statement follows from the Proposition \ref{prop2}.

\item Let $f\in C(\overline\D)$ and $(\mathcal F_\lambda\circ \iota\circ c)(f)$ belongs to $L^0(\H)$ and $\mathcal P^{++}_{pol}$
then its spectrum is discrete and equal to $f(\overline\D)$ at the same time, which leads to a contradiction.

\item From $iii)$ one has that $\proj{\phi}{\psi}\in \mathcal M^{++}_{pol}$ for $\phi,\psi\in\H$. Additionally one has
\be \proj\phi\psi = (\proj uv)(\proj v\phi)\ee
\be \proj\phi v= [\proj \phi\eta, \proj\eta v]\ee
if $v,\eta\in \H$ satisfy $\sc vv=\sc\eta\eta=1$ and $\sc v\eta=0$. Hence $L^0(\H)\subset Comm \mathcal M^{++}_{pol}$.

In order to show that $L^0(\H)\subsetneq Comm \mathcal M^{++}_{pol}$ we observe
that the operator $[a_{11}^\dag,a_{11}]\in L^0(\H)\subset Comm \mathcal M^{++}_{pol}$ in the basis \eqref{baza} assumes the form
\be[a_{11}^\dag,a_{11}]\left|\begin{array}{@{}cc@{}}
  j & m \\
  j_{1}& j_{2} \\
\end{array}\right\rangle=\ee
$$\!\!\!\!\!\!\!\!\!\!\!\!\!\!\!\!\!\!\!\!\!\!\!\!\!\!\!\!\!\!\!\!\!\!\!\!=\frac{(\lambda-2)((j_1+j_2)(m+2j+\lambda)-(m+2j+\lambda)(m+\lambda-2)-(j+j_1+1)(j+j_2+1))}
{(m+2j+\lambda-1)(m+2j+\lambda)(m+\lambda-2)(m+\lambda-1)}\left|\begin{array}{@{}cc@{}}
  j & m \\
  j_{1}& j_{2} \\
\end{array}\right\rangle$$
Thus it is diagonal and $\frac14 \frac{2-\lambda}{(m+\lambda-2)(m+\lambda-1)}$ is the concentration point of
its spectrum. So, it belongs to $Comm\mathcal M^{++}_{pol}$ and is not compact operator.
\item It follows from the fact that $\mathcal P^{++}_{pol}\subset \mathcal P^{++}$.
\end{enumerate}
} 

Now let us make few remarks about the Toeplitz (holomorphic) representation
of $\mathcal M^{++}$, i.e. the representation in the Hilbert space $L^2\mathcal O(\mathbb D,d\mu_\lambda)$.
One obtains it using the anti-linear monomorphism $I:\H\to L^2\mathcal O(\mathbb D,d\mu_\lambda)$ given
by \eqref{anti-I}:
\be \label{TX}\mathcal T_\lambda(X):=I\circ X\circ I^{-1}: L^2\mathcal O(\mathbb D,d\mu_\lambda)\to L^2\mathcal O(\mathbb D,d\mu_\lambda),\ee
where $X\in\mathcal M^{++}$. In the particular case when $X\in\overline{\P^{++}}$
 one has
\be \mathcal T_\lambda(X)\psi(Z)=\langle X\rangle(Z)\psi(Z).\ee
So, $\mathcal T(\overline{\P^{++}_{pol}})$ is realized by multiplication operators $M_f$, $f\in H^\infty(\mathbb D)$, having a continuous
prolongation to $\overline{\mathbb D}$. Thus, the Toeplitz algebra $\mathcal T_\lambda(\mathcal M^{++}_{pol})$ is generated by the operators
\be \label{toeplitz}\mathcal T_\lambda(f)=\Pi_\lambda\circ M_f\circ\Pi_\lambda,\ee
where $f$ is a real analytic polynomial. The operator $M_{f}:L^2\mathcal O(\mathbb D,d\mu_\lambda)\to L^2\mathcal O(\mathbb D,d\mu_\lambda)$ is
the operator of multiplication  by $f\in C(\overline{\mathbb D})$ and
\be (\Pi_\lambda\psi)(Z)=\int_{\mathbb D}\overline{\langle Z|  V\rangle} \psi(V^\dag,V)\frac{1}{\langle V|V\rangle}c_\lambda d\mu(V^\dag,V)\ee
is the orthogonal projector $\Pi_\lambda$ of the Hilbert space $L^2(\mathbb D,d\mu_\lambda)$
on its Hilbert subspace $L^2\mathcal O(\mathbb D,d\mu_\lambda)$.

Using the representation \eqref{TX} one can investigate $\mathcal M^{++}_{pol}$ in the framework
 of theory of Toeplitz algebras related to bounded symmetric domains,
which were intensively investigated in series of works \cite{upmeier1,upmeier2,upmeier}.

The following basic statement can be viewed as a variant of the Coburn Theorem (see \cite{douglas}).
\begin{thm}
One has the exact sequence
\be \label{exact}0\tto Comm\mathcal M^{++}_{pol}\xrightarrow{\quad\iota\quad} \mathcal M^{++}_{pol} \xrightarrow{\quad\pi_\lambda\quad} C(\mathbb M^{00})\tto 0\ee
of $C^*$-algebra homomorphisms, where $C(\mathbb M^{00})$ is the $C^*$-algebra of continuous functions on the conformally
compactified Minkowski space $\mathbb M^{00}$. \end{thm}
\prf{We begin observing that for $f\in C(\overline\D)$ one has inequalities
\be \label{norm-ineq}\norm{\mathcal T_\lambda(f)}_\infty\leq \norm f_{sup}\leq\norm{Q_\lambda(f)}_\infty\ee
which follow from \eqref{toeplitz} and from \eqref{Qf} respectively. From the first inequality in \eqref{norm-ineq} it
follows that the map
\be \label{toep-comm}C(\overline\D)\ni f\tto T_\lambda(f):=[\mathcal T_\lambda(f)]\in \mathcal M^{++}_{pol}/Comm \mathcal M^{++}_{pol}\ee
is a continuous epimorphism of the $C^*$-algebra $C(\overline\D)$ on the commutative quotient $C^*$-algebra $\mathcal M^{++}_{pol}/Comm \mathcal M^{++}_{pol}$.
Let us recall that the norm of $[x]\in\mathcal M^{++}_{pol}/Comm \mathcal M^{++}_{pol}$ is defined by
\be \norm{[x]}_{inf}=\inf_{\xi\in Comm\mathcal M^{++}_{pol}}\norm{x+\xi}.\ee
Now let us consider the ideal $\ker T_\lambda\subset C(\overline\D)$. It follows from $iv)$ of Proposition \ref{prop41} that
$U_\lambda(g)(Comm\mathcal M^{++}_{pol}) U_\lambda(g)^\dag\subset Comm \mathcal M^{++}_{pol}$, so the conformal group
$SU(2,2)/\Z_4$ acts on the quotient $C^*$-algebra $\mathcal M^{++}_{pol}/Comm \mathcal M^{++}_{pol}$ and the $C^*$-algebra
epimorphism defined by \eqref{toep-comm} is a conformally equivariant map, i.e.
\be\label{Ab}\bfig \square<900,500>[C(\overline\D)`\mathcal M^{++}_{pol}/Comm \mathcal M^{++}_{pol}`C(\overline\D)`\mathcal M^{++}_{pol}/Comm \mathcal M^{++}_{pol};T_\lambda`\Sigma_g`{[U_{\lambda}(g)]}`T_\lambda] \efig \ee
for any $g\in SU(2,2)/\Z_4$, where
\be \label{act-S}(\Sigma_gf)(Z^\dag,Z):=f((\sigma_g(Z))^\dag,\sigma_g(Z))\ee
\be [U_\lambda(g)]([x]):=[U_\lambda(g)xU_\lambda(g)^\dag].\ee

We conclude from the above that $\ker T_\lambda$ is an ideal in $C(\overline\D)$ conformally invariant with respect to the action
\eqref{act-S}. Since any ideal in $C(\overline\D)$ consists of functions vanishing on some compact subset $K\subset \overline\D$
the conformally invariant ideals correspond to the conformally invariant compact subsets: $\overline \D$,
$\partial\D=\{Z\in Mat_{2\times2}(\C) : \det(E-Z^\dag Z)=0\textrm{ and }\Tr(E-Z^\dag Z)\geq 0\}$ and $U(2)=\{Z\in Mat_{2\times2}(\C): Z^\dag Z=E\}$, where
the last one is the \v Silov boundary of $\D$.
In this way we show that $\ker T_\lambda$ is equal to one of the following three ideals
\be \mathcal I_{\overline\D}=\{0\}\subset \mathcal I_{\partial\overline\D}\subset\mathcal I_{U(2)},\ee
where by $\mathcal I_{K}$ we denote the ideal of functions equal to zero on $K$.
The polynomial
\be \phi(Z^\dag,Z):=\Tr(E-Z^\dag Z)\ee
generates the ideal $\mathcal I_{U(2)}$ and maps $\overline \D$ on the interval $[0,2]$. Let us consider the positive
operator
\be :\Tr(E-A^\dag A): = 2-a_{11}^\dag a_{11}-a_{12}^\dag a_{12} - a_{21}^\dag a_{21} - a_{22}^\dag a_{22},\ee
which is diagonal, with
\be \label{tr-diag}:\Tr(E-A^\dag A):  \left|\begin{array}{@{}cc@{}}
  j & m \\
  j_{1}& j_{2} \\
\end{array}\right\rangle=\frac{2(\lambda-2)(m+j+\lambda-1)}{(m+\lambda-1)(m+2j+\lambda)}\left|\begin{array}{@{}cc@{}}
  j & m \\
  j_{1}& j_{2} \\
\end{array}\right\rangle,\ee
in the basis \eqref{baza}. We see from \eqref{tr-diag} that the spectrum $\sigma$ of $:\Tr(E-A^\dag A):$ is contained
in the interval $[0,2]$ and the set
\be \sigma_a:=\left\{\frac{\lambda-2}{m+\lambda-1}: m\in\No\cup\{\infty\}\right\}\ee
 is its approximative spectrum.
The continuous function $F:[0,2]\to \R$ defined by
\be F(x):=x\sin\frac{(\lambda-2)\pi}{x}\ee
vanishes on $\sigma_a$ and $F\circ\phi\in \mathcal I_{U(2)}$. Since $F_{|\sigma_a}\equiv 0$ and $F$ assumes the same value
at most on a finite subset of $\sigma\setminus\sigma_a$, we conclude that $F(:\Tr(E-A^\dag A):)$ is a compact operator. Thus, by $iii)$ of Proposition
\ref{prop41} $F(:\Tr(E-A^\dag A):)$ belongs to $Comm\mathcal M^{++}_{pol}$. Let us take the sequence $\{P_n(x)\}_{n\in\N}$ of polynomials
which uniformly approximate $P_n\to F$ the function $F\in C([0,2])$. Thus one has
\be \label{F-approx}\norm{P_n\circ\phi-F\circ\phi}_{sup}\xrightarrow[n\to\infty]{} 0\ee
From \eqref{F-approx} and the first inequality of \eqref{norm-ineq} we obtain
\be \label{TF}\norm{\mathcal T_\lambda (P_n\circ\phi)-\T_\lambda(F\circ\phi)}_\infty\xrightarrow[n\to\infty]{} 0 .\ee
On the other hand, from the Gelfand-Naimark theorem and \eqref{F-approx} we have
\be \label{P-Tr}\norm{P_n(:\Tr(E-A^\dag A):)-F(:\Tr(E-A^\dag A):)}_\infty\xrightarrow[n\to\infty]{} 0 .\ee
The operators $\mathcal T_\lambda(P_n\circ\phi)$ are polynomials of the creation and annihilation operators taken in the
anti-normal ordering, so they differ from the polynomials $P_n(:\Tr(E-A^\dag A):)$ modulo elements of $Comm\mathcal M^{++}_{pol}$.
Thus, using also \eqref{TF} and \eqref{P-Tr}, we obtain that
\be 0=\norm{[\mathcal T_\lambda(F\circ\phi)]-[F(:\Tr(E-A^\dag A):)]}_{inf}=\norm{[\mathcal T_\lambda(F\circ\phi)]}_{inf}=
\norm{T_\lambda(F\circ\phi)}_{inf}.\ee

Summing up we conclude that $F\circ \phi\in \ker T_\lambda \cap \mathcal I_{U(2)}$. Since it is easy to check that $F\circ\phi\notin \mathcal I_{\partial\D}$
and that $\ker T_\lambda$,  is conformally invariant it follows that $\ker T_\lambda=\mathcal I_{U(2)}=\mathcal I_{\mathbb M^{00}}$.

Taking into account that \eqref{toep-comm} is an epimorphism of $C^*$-algebras, we state the following
isomorphisms  $\mathcal M^{++}_{pol}/Comm \mathcal M^{++}_{pol}\cong C(\overline\D)/\mathcal I_{\mathbb M^{00}}\cong C(\mathbb M^{00})$.
These isomorphisms give the epimorphism $\pi_\lambda:\mathcal M^{++}_{pol}\to C(\mathbb M^{00})$.
} 

Ending this section, let us remark that ''neglecting'' the non-commutativity of quantum complex Minkowski space $\mathcal M^{++}_{pol}$
we come back to the commutative $C^*$-algebra $C(\mathbb M^{00})$ whose spectrum is given by the conformally compactified Minkowski space $\mathbb M^{00}$.

\section{Quantization and physical interpretation}

Analogously to the classical coordinate observables $(Z,Z^\dag)$ on $\mathbb M^{++}$ we shall use quantum coordinate
observables $(\mathbb A,\mathbb A^\dag)$ for the quantum phase space $\mathcal M^{++}$. Superposing morphisms from diagram \eqref{diagram} we obtain the extension of this correspondence.
In such a way we get the
isomorphism
\be \label{quant-iso}Q_\lambda:=\mathcal F_\lambda \circ \iota\circ c:\mathcal B(\D)\tto\Li,\ee
which extends the quantization map,
\be a:H^\infty(\D)\ni f\tto a(f)\in\Li,\ee
 discussed in the previous section. Taking into account the properties
\be Q_\lambda(f*_\lambda g)=Q_\lambda(f)Q_\lambda(g),\ee
\be Q_\lambda(\bar f)=Q_\lambda(f)^*,\ee
\be \label{Qf}\mn{Q_\lambda(f)}_\lambda=f,\ee
for $f,g\in\mathcal B(\D)$, we see that the isomorphism $Q_\lambda$ gives a \emph{quantization procedure} inverse to the mean value map.

According to relation \eqref{Qf}, Berezin covariant symbols are the classical observables
corresponding to the quantum observables realized by the bounded operators. As a particular case the quantum
phase space $\mathcal M^{++}\subset\Li$ is obtained from $\mn{\mathcal M^{++}}\subset \mathcal B(\D)$
by the quantization \eqref{quant-iso}. However for physical reasons we are interested in the
extension of $Q_\lambda:\mathcal B(\D)\to\Li$ to a larger algebra of observables. For example it is reasonable to include in this
scheme the elements of the enveloping algebra of the conformal Lie algebra $su(2,2)$. The latter ones are represented by
unbounded operators in $\H$ which, according to the equivariance property \eqref{Ab}, possess the common domain given by
the linear span $\mathcal L(\mathcal K_\lambda(\M^{++}))$ of the set $\mathcal K_\lambda(\M^{++})$ of the coherent states.
Let us then define the vector space $\mathcal A^{++}$ of operators in $\H$ closed with respect to the
operation of conjugation and all elements of which possess $\mathcal L(\mathcal K_\lambda(\M^{++}))$  as a common domain. Therefore
for any operator
$F\in\mathcal A^{++}$ the 2-covariant and Berezin covariant symbols have sense.

In the following we will use the coherent state weak topology, i.e. $A_n\xrightarrow{coh} A$ if
$\sc{Z|A_n}{V}\to\sc{Z|A}V$ for all $Z,V\in\D$. It is a weaker topology than the weak one, as can be seen from the following example.
Let $\D\ni Z_n=(1-\frac1n)E$, $n\in\N$. We define the sequence of operators
\be A_n:=n\frac{\proj{Z_n}{Z_n}}{\sc{Z_n}{Z_n}}.\ee
It is easily observed that
\be \forall Z,V\in \D\quad \lim_{n\to\infty} \sc{Z|A_n}{V}=0, \ee
thus $A_n\xrightarrow{coh} 0$. On the other hand $\sup_{n\in\N}\norm{A_n}=\infty$, thus $A_n$ is not weakly convergent.

The space $\mathcal A^{++}$ is closed with respect to coherent state weak topology. The quantum phase space $\mathcal M^{++}$
is contained in $\mathcal A^{++}$ as a dense subset with respect to the coherent state weak topology. For any $F\in\mathcal A^{++}$
its Berezin symbol $f=\mn F\in\mathcal R\O^{++}(\D)$ is the real analytic function
\be f(Z^\dag,Z)=\sum f_{i_{11},i_{12},i_{21},i_{22},j_{11},j_{12},j_{21},j_{22}}\bar Z_{11}^{i_{11}}\bar Z_{12}^{i_{12}}
\bar Z_{21}^{i_{21}}\bar Z_{22}^{i_{22}} Z_{11}^{j_{11}}Z_{12}^{j_{12}}Z_{21}^{j_{21}}Z_{22}^{j_{22}}\ee
of the variables $(Z^\dag,Z)$. One extends the quantization \eqref{quant-iso} naturally to the space $\mathcal R\O^{++}(\D)$
of real analytic functions on $\D$ by setting
\begin{align} \label{Qff} Q_\lambda(f)=&\sum f_{i_{11},i_{12},i_{21},i_{22},j_{11},j_{12},j_{21},j_{22}} {a_{11}^\dag}^{i_{11}}
{a_{12}^\dag}^{i_{12}}{a_{21}^\dag}^{i_{21}}{a_{22}^\dag}^{i_{22}}a_{11}^{i_{11}}a_{12}^{i_{12}}a_{21}^{i_{21}}a_{22}^{i_{22}}= \nonumber\\
=&\; :f(\mathbb A^\dag,\mathbb A):,\end{align}
where as usual, the colons $:\cdot:$ denote normal ordering. The infinite sum in \eqref{Qff} is taken in
the sense of coherent state weak topology. The extension of the product $*_\lambda$, see \eqref{star}, to the real analytic Berezin symbols
$f,g\in\mathcal R\O^{++}(\D)$ is defined by
\be (f *_\lambda g)(Z^\dag,Z):=\frac{\sc{Z^\dag|:f(\mathbb A^\dag,\mathbb A)::g(\mathbb A^\dag,\mathbb A):}{Z}}{\sc{Z^\dag}{Z}}.\ee
As an illustration let us consider the Berezin symbols
\be \mn{U_\lambda(g)}(Z^\dag,Z)=(\det(CZ+D))^{-\lambda}\left(\frac{\det(E-Z^\dag\sigma_g(Z))}{\det(E-Z^\dag Z)}\right)^{-\lambda}\ee
and their quantum $(\mathbb A^\dag,\mathbb A)$-coordinate representation
\be \label{Ulamb}U_\lambda(g)=Q_\lambda(\mn{U_\lambda(g)})=:\left(\frac{\det(E-\mathbb A^\dag\sigma_g(\mathbb A))}{\det(E-\mathbb A^\dag \mathbb A)}\right)^{-\lambda}:(\det(C\mathbb A+D))^{-\lambda}\ee
for the conformal group elements $g\in \rm{SU}(2,2)$. In order to express the quantum 4-momentum, relativistic angular momentum, dilation
and 4-acceleration in terms of quantum coordinates $(\mathbb A^\dag,\mathbb A)$ we  differentiate $U_\lambda(g(t))$ given by
\eqref{Ulamb} with respect to the parameter $t\in\R$ for an appropriate choice of one-parameter subgroup $\R\ni t\to g(t)\in \rm{SU}(2,2)$.
As a result one obtains
\begin{align}
\label{Qp}Q_\lambda(p_\mu)&=i\lambda : (\det(\mathbb W- \mathbb W^\dag))^{-1}\Tr(\sigma_\mu(\mathbb W-\mathbb W^\dag)):\\
Q_\lambda(m_{\mu\nu})&=i\lambda\left(\frac{1}{2}\Tr(\sigma_\mu \mathbb W^\dag) : (\det(\mathbb W- \mathbb W^\dag))^{-1}\Tr(\sigma_\nu(\mathbb W-\mathbb W^\dag)): -\nonumber \right.\\
    &\left.- \frac{1}{2}\Tr(\sigma_\nu \mathbb W^\dag): (\det(\mathbb W- \mathbb W^\dag))^{-1}\Tr(\sigma_\mu(\mathbb W-\mathbb W^\dag)):\right)\\
Q_\lambda(d)&=i\lambda \Tr(\sigma_\mu \mathbb W^\dag) : (\det(\mathbb W- \mathbb W^\dag))^{-1}\Tr(\sigma^\mu(\mathbb W-\mathbb W^\dag)): - 2i\lambda \mathbb{I}\\
\label{Qa}Q_\lambda(a_\nu)&=i\lambda\det(\mathbb W^\dag) : (\det(\mathbb W- \mathbb W^\dag))^{-1}\Tr(\sigma_\nu(\mathbb W-\mathbb W^\dag)): -\nonumber\\
 &- i\lambda \frac{1}{2}\Tr(\sigma_\nu \mathbb W^\dag)\Tr(\sigma^\beta \mathbb W^\dag)  : (\det(\mathbb W- \mathbb W^\dag))^{-1}\Tr(\sigma_\beta(\mathbb W-\mathbb W^\dag)):+\nonumber\\
  &+ i\lambda\Tr(\sigma_\nu \mathbb W^\dag),
\end{align}
where $(\mathbb W^\dag,\mathbb W)$ are matrix operator coordinates in $\mathcal A^{++}$ obtained from $(\mathbb A^\dag,\mathbb A)$ by the
Caley transform
\be \mathbb W=i(\mathbb A+E)(\mathbb A-E)^{-1},\ee
which has sense in the coherent state weak topology. After passing to the representation in the Hilbert space $L^2\O(\mathbb T,d\mu_\mu)$
of holomorphic functions
on the future tube $\mathbb T$, square integrable with respect to the measure \eqref{miara}, we rediscover from \eqref{Qp}-\eqref{Qa} the operators \eqref{hat-p}-\eqref{hat-a}
obtained by the Kostant-Souriau geometric quantization.

It follows from \eqref{hat-p} that
\be \label{Qp-comm}[Q_\lambda(p_\mu),Q_\lambda(p_\nu)]=0.\ee
Using \eqref{pp}, we see from \eqref{Qp-comm} that
\be [Q_\lambda(y_\mu),Q_\lambda(y_\nu)]=0.\ee
The creation operators
\be \label{cre-op}Q_\lambda(\bar w^\mu)=\frac12\Tr(\sigma_\mu \mathbb W^\dag)\ee
in $L^2\O(\mathbb T,d\mu_\lambda)$ are given as multiplication by the complex coordinate functions $w^\mu$, so they commute.
Thus, in  addition to \eqref{Qp-comm}, we have
\be [Q_\lambda(x^\mu),Q_\lambda(x^\nu)]=0\ee
\be\label{CCR} [Q_\lambda(x^\mu),Q_\lambda(p_\nu)]=-i\delta^\mu_\nu 1\ee
for the quantum canonical coordinates $(Q_\lambda(x^\mu),Q_\lambda(p_\nu))$.

Therefore we see that Heisenberg algebra generated by unbounded operators of 4-momenta $Q_\lambda(p_\nu)$ and 4-positions
$Q_\lambda(x^\mu)=\frac12\Tr(\sigma_\mu(\mathbb W+\mathbb W^\dag))$ is included  in $\mathcal A^{++}$. The creation
operators \eqref{cre-op} and the annihilation ones
\be Q_\lambda(w_\nu)=\frac12\Tr(\sigma_\nu\mathbb W)\ee
generate the Caley transforms of quantum polarizations $\overline{\mathcal P^{++}_{pol}}$ and $\mathcal P^{++}_{pol}$ respectively.
However their commutators $[Q_\lambda(\bar w^\mu),Q_\lambda(w_\nu)]\neq 0$ do not have so simple form as it  has place in the case of quantum
real polarization given by the canonical commutation relation \eqref{CCR}.

Let us now discuss the physical sense of the parameter $\lambda\in\R$. So far, for technical reasons, we
assumed that it was dimensionless. However, as one sees from \eqref{pp}, $\lambda$ has dimensions of action.
We therefore assume the Planck constant $h$ as the natural unit for $\lambda$. After this we obtain
\be w^\mu=x^\mu+i\lambda \frac{h}{mc}\frac{p^\mu}{mc},\ee
where $mc=\sqrt{p_0^2-\vec p^2}$. The quantity $\frac{h}{mc}$ is the Compton wavelength of the conformal particle. For example
for the proton $\frac{h}{mc}\cong 10^{-13}cm$.

The quantities $\frac{p^\mu}{mc}$ denote the components of relativistic 4-velocity measured
with the speed of light as the unit. Dimensional analysis shows
that in the limit $\lambda\to\infty$ the theory describes physical phenomena characterized by a space-time scale much bigger than
the Compton scale characteristic for the quantum phenomena. This physical argument is consistent with the following asymptotic
behavior of of $*_\lambda$-product
\be \label{star-mult}f*_\lambda g\sim fg\ee
\be \label{star-pois}f*_\lambda g-g*_\lambda f\sim i\lambda\{f,g\}\ee
for $\lambda\to\infty$, where the right hand side of \eqref{star-mult} is usual multiplication of functions and
the right side of \eqref{star-pois} is the Poisson bracket \eqref{poiss}.
In order to show these asymptotic formulae we apply the method used for the case of a general symmetric domain in \cite{ber1}.
The expressions \eqref{star-mult}, \eqref{star-pois} show the correspondences of the
quantum description of the massive scalar conformal particle to its classical mechanical description in the large space-time scale limit.

The quantum effects are described by the transition amplitude \eqref{alamb}, which in the coordinates $(\bar w^\mu,w^\nu)$
is given by
\be \label{amp-coord}a_\lambda(v^\dag,w)=\left(\frac{((w-\bar w)^2(v-\bar v)^2)^\frac12}{(w-\bar v)^2}\right)^\lambda,\ee
where $(w-\bar v)^2=\eta_{\mu\nu}(w^\mu-\bar v^\mu)(w^\nu-\bar v^\nu)$ and $\lambda >3$. One sees from \eqref{amp-coord} that
the transition probability $\abs{a_\lambda(v^\dag,w)}^2$ from $w$ to $v$ as a function of $v$ forms a narrow peak around the coherent
state $w\in\mathbb T$ if $\lambda\frac{h}{mc}\approx 0$.
A more detailed physical discussion can be found in \cite{Oker}.

\section*{Acknowledgement}

Authors thank Tomasz Goli\'nski and Ian Marshall for interest in the paper and corrections to the manuscript.

\end{document}